\documentclass[journal=jacsat,manuscript=article]{achemso}
\usepackage{achemso}
\usepackage{comment}
\usepackage[version=3]{mhchem} 

\author{Omri Meron}
    \affiliation[CMD]{Condensed Matter Physics Department, School of Physics and Astronomy, Faculty of Exact Sciences, Tel Aviv University, Tel Aviv 6997801, Israel}
    \alsoaffiliation[LMI]{Center for Light‐Matter Interaction, Tel Aviv University, Tel Aviv 6997801, Israel}
\email{omrimeron@tauex.tau.ac.il}

\author{Snir Nehemia}
    \affiliation[CMD]{Condensed Matter Physics Department, School of Physics and Astronomy, Faculty of Exact Sciences, Tel Aviv University, Tel Aviv 6997801, Israel}
    \alsoaffiliation[LMI]{Center for Light‐Matter Interaction, Tel Aviv University, Tel Aviv 6997801, Israel}

\author{Uri Arieli}
    \affiliation[CMD]{Condensed Matter Physics Department, School of Physics and Astronomy, Faculty of Exact Sciences, Tel Aviv University, Tel Aviv 6997801, Israel}
    \alsoaffiliation[LMI]{Center for Light‐Matter Interaction, Tel Aviv University, Tel Aviv 6997801, Israel}

\author{Haim Suchowski}
    \affiliation[CMD]{Condensed Matter Physics Department, School of Physics and Astronomy, Faculty of Exact Sciences, Tel Aviv University, Tel Aviv 6997801, Israel}
    \alsoaffiliation[LMI]{Center for Light‐Matter Interaction, Tel Aviv University, Tel Aviv 6997801, Israel}

\title[]
  {Shaping Ultrafast Pulses for Enhanced Resonant Nonlinear Interactions}

\abbreviations{}

\begin{document}


\begin{abstract}
Coherent control with shaped ultrafast pulses is a powerful approach for steering nonlinear light–matter interactions. Previous studies in quantum control have shown that, beyond transform-limited pulses, those with antisymmetric spectral phases can drive nonresonant multiphoton transitions with comparable efficiency. However, in resonant multiphoton transitions, the material's spectral-phase response introduces dispersion that degrades nonlinear efficiency. Pre-shaping the pulse to compensate for the material's impulse response can restore and enhance nonlinear interactions beyond the transform-limited case. Yet, is this the only spectral phase that can yield such enhancement? Here, we study sub-10 fs single-pulse four-wave mixing in resonant plasmonic nanostructures using arctangent spectral-phase-shaped pulses. We uncover two distinct enhancement regimes: one compensating for material dispersion, and a counterintuitive regime where the arctangent phase induces an antisymmetric polarization response, driving constructive multiphoton pathway interference. Our theoretical analysis provides clear physical explanation for both phenomena. Notably, it predicts that both enhancement mechanisms scale exponentially with harmonic order, offering a powerful strategy for dramatically enhancing high-order harmonic generation in resonant systems.
\end{abstract}

\section{Introduction }

Coherent control using shaped ultrafast optical pulses has emerged as a powerful framework for steering quantum and nonlinear interactions in matter.\cite{warren1993} By tailoring the spectral phase, amplitude, and polarization of femtosecond pulses,\cite{weiner2011} one can selectively manipulate excitation pathways with exceptional precision.\cite{assion1998, meshulach1998, bartels2000, levis2001, aeschlimann2007, piatkowski2016, remesh2018, kappe2025}  Among these parameters, spectral phase plays a particularly central role, especially in resonant multiphoton processes, where interactions are inherently non-instantaneous and strongly influenced by the system’s spectral response.

A pivotal advance was made by the Silberberg group, who introduced the concept of quantum control of multiphoton transitions using shaped pulses in both nonresonant and resonant atomic media.\cite{silberberg2009} In their seminal work, they showed that a pulse with a spectral phase antisymmetric around half the two-photon resonance frequency could efficiently drive multiphoton excitations, achieving results comparable to transform-limited (TL) pulses.\cite{meshulach1998, meshulach1999} This finding was strikingly counterintuitive: although such shaped pulses can exhibit extremely long temporal durations and very low peak power, they nonetheless produce maximal nonlinear responses due to the constructive interference of multiphoton pathways. Further, Silberberg and colleagues demonstrated that TL pulses are often not optimal for resonant multiphoton interactions, highlighting the need to tailor the spectral amplitude or phase to compensate for the system’s resonant response.\cite{dudovich2001} In particular, they showed that applying a $\pi$-step spectral phase centered at the resonance frequency, shaping the incoming pulse to counteract the resonant spectral phase, enables temporal compression that leads to a transient enhancement of the induced transitions.\cite{dudovich2002a} These foundational insights established spectral phase shaping as a critical strategy for enhancing nonlinear interaction strengths and revealed the deep connection between phase symmetry, temporal dynamics, and quantum interference in multiphoton processes.

Originally demonstrated in isolated atomic and molecular systems, pulse shaping has since evolved into a versatile technique applied across diverse platforms, including nanostructures and condensed-matter systems.\cite{shcherbakov2019, bahar2020a, accanto2021, giegold2022, lange2024, farhi2025} Stockman proposed extending these concepts to nanoplasmonics, showing that the temporal profile of excitation pulses could be tailored to match the time-reversed resonant dynamics of carriers in metallic nanoparticles.\cite{stockman2002, stockman2008} This strategy, aimed at localizing optical energy far below the diffraction limit, led to various theoretical and experimental demonstrations.\cite{brixner2004,utikal2010, piatkowski2016, stockman2018} Notably, Huang introduced a deterministic framework in which spectral phase functions derived from FDTD simulations were used to compress plasmonic responses in time, enabling control over ultrafast nanoscale fields.\cite{huang2009} Recent experiments have reaffirmed the critical role of spectral phase in such regimes. For example, Bahar et al. demonstrated coherent control of second-order nonlinear enhanced emission in U-shape plasmonic resonance by optimizing chirped pulses, clearly revealing the non-instantaneous character of the interaction.\cite{bahar2022} 

While controlling resonant multiphoton interactions in complex systems remains challenging due to rapid decoherence and many-body effects, in our recent study on coherent control of two-dimensional semiconductors, we further advanced these ideas by demonstrating that in order to maximize the third-order nonlinear response near resonance, we introduced an arctangent (Atan) spectral phase function, precisely matched in center and width to the exciton resonance frequency and its decoherence rate. This function effectively counteracts the dispersion induced by the interaction of the 2D excitonic resonance, enabling a compressed temporal polarization response that enhances nonlinear optical performance. Importantly, this method allowed simultaneous compensation of multiple resonances within the pulse bandwidth, showcasing the strength of spectral phase-based coherent control in complex materials.\cite{meron2025}

Here, we systematically map the Atan spectral phase space of the ultrafast resonant plasmonic response by scanning its central frequency and spectral width. We employ shaped single-pulse, four-wave mixing (FWM) to probe $\chi^{(3)}$ processes. By mapping the nonlinear multiphoton enhancement and suppression landscape across the spectral phase space, we uncover a symmetric structure governed by detuning and phase parity. 
This phase-space topology reveals a well-established enhancement region where the spectral phase naturally compensates for the plasmonic resonant dephasing,\cite{huang2009, bahar2022, meron2025} allowing for direct extraction of the near-field resonant frequency and linewidth. More surprisingly, we identify a second, previously unexplored enhancement region that arises from antisymmetric phase relations, where the applied phase adds dispersion rather than compensates for the resonance-induced spectral phase, creating constructive multiphoton interference through a fundamentally different mechanism. This finding bridges insights from both resonant and nonresonant coherent control schemes, offering a unified framework for exploiting constructive multiphoton interference in complex media. To support these findings, we develop a compact second-order model that captures the observed enhancement patterns and show the approach extends to higher-order processes, providing significant enhancement factors for perturbative high-harmonic generation. Our results demonstrate that spectral phase symmetry and detuning fundamentally govern multiphoton efficiency in resonant systems, enabling rational design of shaped pulses for enhanced nonlinear processes. 
 
\section{Results and discussion}

\begin{figure} 
        \centering
        \includegraphics[width=1\textwidth]{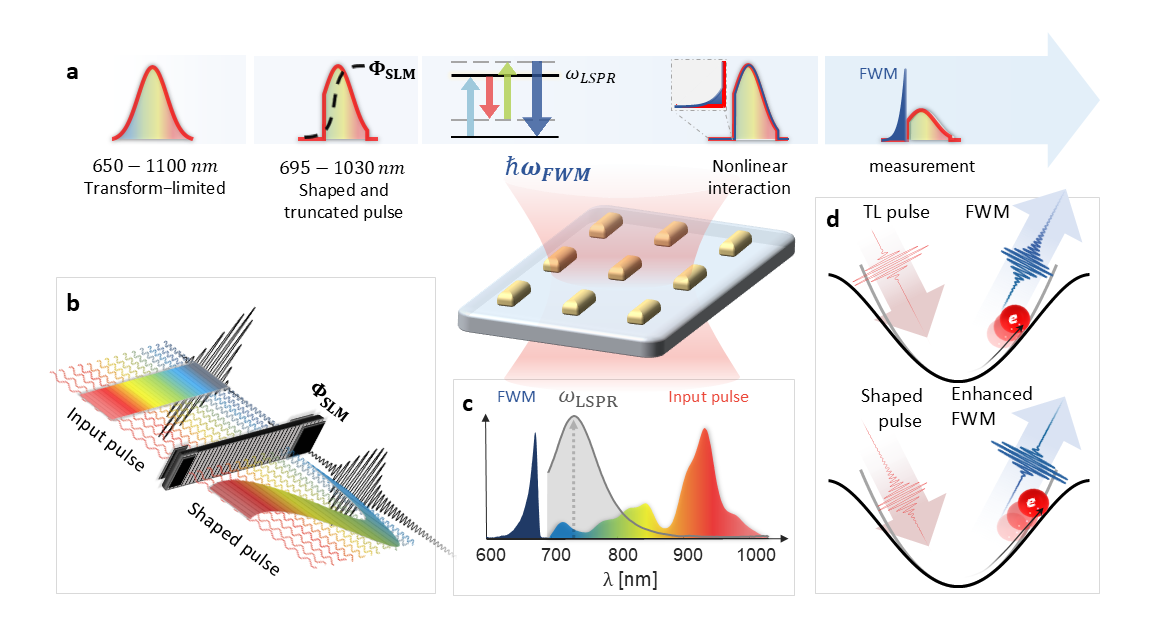}
        \caption{
       (a) A diagram of the single-pulse FWM experimental apparatus in the frequency domain: a sub-10\,fs pulse is shaped and spectrally truncated by the SLM (see (b)). The pulse is then focused onto an array of gold nanobars, generating a FWM nonlinear response that is detected by a spectrometer via reflection.
(b) Schematic illustration of a negative Atan spectral phase applied by the SLM, and how it modifies the wavelength-dependent group-delay. 
(c) Power spectral density (colorful area) of the ultrabroadband sub-10\,fs input pulse used in the experiment, overlaid with the linear reflection spectrum of the gold nanobar array (gray), acquired using the same excitation pulse. The resulting FWM signal (dark blue), collected in reflection and scaled for visibility, appears on the anti-Stokes side of the spectrum.
(d) Illustration of the quartic correction to the harmonic potential, giving rise to third-order nonlinearity from the LSPR. A shaped driving pulse can increase the transient oscillator displacement, optimizing nonlinearity compared to a TL pulse.}
\label{fig1}
\end{figure}

In our experiments, we utilize a spatial light modulator (SLM)-based pulse shaping setup, as depicted in Fig. \ref{fig1}. The SLM is positioned in the Fourier plane of a 4f system \cite{weiner2011}. This arrangement spatially disperses the sub-10\,fs ultrabroadband pulse into aligned spectral components. The SLM enables precise temporal shaping of the spectral phase, $\phi_{SLM}(\lambda)$, as illustrated in Fig. \ref{fig1}.b. Additionally, the Fourier plane serves as a sharp edge filter which helps truncate the blue end of the spectrum. Upon passing through a tightly focusing mirror objective (Pike, NA-0.78), the shaped and truncated pulse interacts with an array of gold nanobars. The pulse is linearly polarized along the nanobars' elongated axis.  The reflection spectrum of the nanobars, as measured with our ultrabroadband pulse (gray shaded area in Fig. \ref{fig1}.c), displays a pronounced resonance peak at $\omega_{LSPR} = 1.68 \text{ eV} = 738 \text{ nm}$, verifying the effective photo-excitation of the localized surface plasmon resonance (LSPR) mode. Importantly, the spectral bandwidth of the pulse is substantially broader than the LSPR linewidth, facilitating the simultaneous excitation across the entire bandwidth of the metallic nanostructure's LSPR response. This is crucial for achieving coherent control over the excitation.  

Beyond linear interactions, the ultrabroadband pulse also drives various intrapulse nonlinear wave-mixing processes. Due to interband absorption and the centrosymmetric geometry of the gold nanobar, sum-frequency generation (SFG) was not observed. Instead, we focus our measurements on the FWM signal, which arises from the coherent nonlinear interaction of three frequency components within the pulse, combining along distinct optical pathways with well-defined phase relationships.  This FWM signal is collected in reflection on the anti-Stokes side of the pulse, outside the truncated pulse spectrum, with the fundamental attenuated using a short-pass edge filter. The experimental apparatus and FWM detection scheme are depicted in Fig.\ref{fig1}.\cite{dudovich2002, suchowski2013, meron2025, kravtsov2016}

\begin{figure} 
        \centering
        \includegraphics[width=1\textwidth]{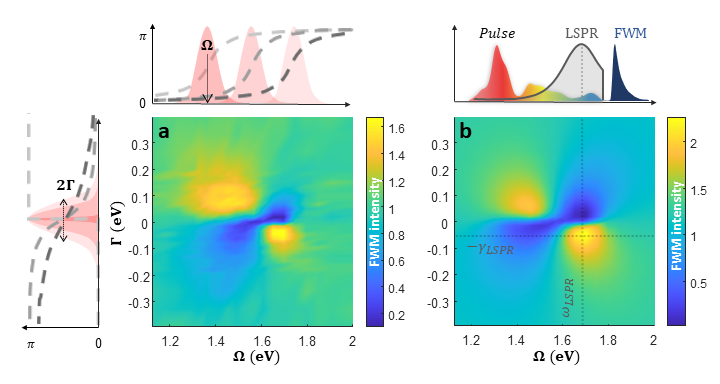}
        \caption{Measured and simulated 2D landscapes of the Atan phase-space scan.
(a) A measurement of the integrated FWM intensity (normalized to the TL case) as a function of the Atan spectral phase parameters: linewidth $\Gamma$ and central frequency $\Omega$, as defined in Eq.~\ref{eq0}. Illustration of an Atan phase, applied by the SLM, while varying  $\Gamma$ for a given $\Omega$ (left panel) and vice versa (top panel). At the edges of the measurement, the phase approaches the TL case. 
(b) Simulated 2D phase-space map of the integrated FWM response based on the AHO model (Eq.~\ref{eq3}). The top panel includes the spectral bandwidth of the input pulse, the position of the LSPR, and the measured FWM signal, all plotted on the same energy scale as the $\Omega$ axis.} 
        \label{fig2}
        \end{figure}

We use the SLM to systematically scan the full parameter space of the Atan phase function, scanning both the central frequency $\Omega$ and linewidth $\Gamma$:
\begin{equation}
\phi_E(\omega) = \tan^{-1}\left(\frac{2\Gamma\omega}{\Omega^{2} - \omega^{2}}\right).
\label{eq0}
\end{equation}
This parametrization allows us to tailor the spectral phase to either compensate or add dispersion to the intrinsic resonant phase of the system. As shown in Fig.~\ref{fig2}a, the obtained two-dimensional FWM intensity map reveals a rich landscape of multi-photon pathway interference responses. First, we observe the primary enhancement FWM region in Fig.~\ref{fig2}a, which closely corresponds in central frequency to the LSPR $\Omega=\omega_{{LSPR}} = 1.68\,eV$, whereas the optimal linewidth, $\Gamma=-\gamma_{LSPR} = 0.049\pm0.015\,eV$, is notably narrower (in absolute value) than the linewidth measured in linear reflection, $\gamma^{lin}_{LSPR} = 0.99\,eV$, using the same ultrabroadband laser source (see supplementary information). 
These results highlight two key differences from far-field linear measurements: (i) the measured FWM signal reflects localized near-field resonant properties that can deviate from spatially averaged far-field linear responses,\cite{zuloaga2011, metzger2016, arieli2019} and (ii) the sub-10\,fs nonlinear response probes the system near its homogeneous broadening limit, effectively filtering out slower inhomogeneous effects.

The measured phase landscape in Fig.~\ref{fig2}a also reveals intriguing secondary regions of enhanced and suppressed nonlinear responses. 
To disentangle the multiphoton interference effects underlying our experimental observations, we model the nonlinear dipolar LSPR response using an anharmonic oscillator (AHO) framework, extending the classical harmonic description of electron displacement by incorporating quadratic, cubic \cite{metzger2016}, and higher-order nonlinearities.\cite{boyd2023}  
The electron displacement $x(t)$, driven by the ultrafast electric field $E(t)$, is characterized by the following equation:
\begin{equation}
\ddot{x}(t) + 2\gamma_0 \dot{x}(t) + \omega_0^2 x(t) + \sum_{n=2}^\infty\alpha_n x^n(t) = -\frac{e E(t)}{m} ,
\label{eq1}
\end{equation}
where $\omega_0$ and $\gamma_0$ are the LSPR resonance frequency and linewidth, respectively, and $\alpha_n$ are the n\textsuperscript{th}
order nonlinear coefficients. Concentrating on third-order nonlinearity ($n=3$) and assuming a perturbative regime, we decompose the displacement as $x(t) = x_0(t) + \delta x(t)$, where $x_0(t)$ is the linear (Lorentz) oscillator solution, and $\delta x(t)$ is the nonlinear correction. In the frequency domain, the linear response is:
\begin{equation}
\tilde{x}_0(\omega) \propto \tilde{E}(\omega) D(\omega) e^{i(\phi_E +\phi_D)},
\label{eq2}
\end{equation}
with $D(\omega) = (\omega_0^2 - \omega^2 - 2i\omega \gamma_0)^{-1}$. Thus, the third-order nonlinear correction will scale as a 3-fold auto-convolution:
\begin{equation}
\delta\tilde{x}^{(3)}(\omega) \approx -\alpha_3 D(\omega)
\big[\tilde{x}_0(\omega) * \tilde{x}_0(\omega) * \tilde{x}_0(\omega)\big].
\label{eq3}
\end{equation}
Differing from narrow-band source treatments, which simplify convolution to a sum over selected discrete mixing frequencies \cite{boyd2023}, we focus on retaining all intra-pulse four-wave interactions within our ultra-broadband source. 

From Eq. (\ref{eq2}), we can observe that the nonlinear polarization $P^{(3)}\propto\lvert \delta \tilde{x}^{(3)}(\omega) \rvert^2$ is maximized when $\tilde{x}_{0}(\omega)=\lvert \tilde{x}_{0}(\omega) \rvert$. To achieve this, the electric field phase can be tailored to counteract the resonant phase in Eq. \ref{eq2} by setting $\phi_E(\omega) = -\phi_D(\omega) =- \tan^{-1}(\frac{2\gamma_{0}\omega}{\omega_{0}^{2}-\omega^{2}})$, leading to an optimally compressed (i.e. TL) oscillator displacement (as illustrated in Fig. \ref{fig1}c).\cite{huang2009, meron2025} Moreover, the Atan phase enhances the instantaneous displacement of the oscillator. As the oscillator is driven away from its minimum energy, the potential becomes less quadratic, thereby increasing the nonlinear polarization (see illustration in Fig.\ref{fig1}.d).

Figure~\ref{fig2}b shows the simulated FWM response from our AHO model as a function of the Atan phase parameters ${\Omega}$ and ${\Gamma}$, using the same driving pulse and LSPR parameters extracted from the experiment.
Remarkably, the simulated FWM response reproduces the experimental landscape with striking accuracy, capturing the primary enhancement peak centered at resonance and partially reflecting the structure of the secondary enhancement and suppression regions. Both experimental and simulated maps display a characteristic four-quadrant pattern, where each quadrant corresponds to a distinct regime of multiphoton pathway interference. However, while the AHO model quantitatively matches the data, the physical intuition behind the secondary enhancement remains elusive. This motivates our turn to a simpler nonlinear model, which offers clearer insight into the symmetry and interference features observed in the phase-space landscape.

\begin{figure} 
        \centering
        \includegraphics[width=1\textwidth]{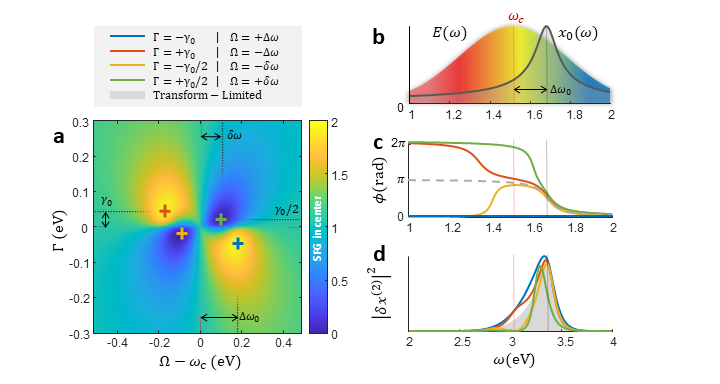}
\caption{Theoretical analysis of multiphoton pathway interference using a second-order nonlinear model.
(a) Simulated SFG intensity at fixed detection frequency $2\omega_c$ as a function of Atan phase center $(\Omega - \omega_c)$ and linewidth $\Gamma$, revealing a characteristic four-quadrant symmetry.
(b) Spectral profiles of the driving Gaussian field $E(\omega)$ and a blue-detuned Lorentzian resonance $x_0(\omega)$.
(c) Spectral phase of the oscillator displacement $x_0(\omega)$ for selected points in panel (a).
(d) Simulated SFG power spectral density $|\delta x^{(2)}(\omega)|^2$ corresponding to the configurations in panel (c), highlighting broadening and narrowing of the spectrum via constructive and destructive multiphoton pathway interference.}
        \label{fig3}
 \end{figure}

We simulate the Atan phase-space response of a second-order nonlinear oscillator  (setting $n=2$ in Eq.~\ref{eq1}), focusing on SFG under a 6\,fs Gaussian driving pulse with carrier frequency $\omega_c$. Figure~\ref{fig3}a displays the resulting nonlinear response at the fixed detection second harmonic frequency $2\omega_c$. Similar to the linewidth $\Gamma$ axis, which is scanned from negative to positive values, effectively reversing the sign of the imposed Atan spectral phase, we set the frequency $\Omega$ axis relative to $\omega_c$. This representation naturally defines a coordinate system centered on two symmetry axes: (i) a vertical line at $\Omega = \omega_c$, which roughly distinguishes between on/off resonance excitation, and (ii) a horizontal line at $\Gamma = 0$, which separates phase functions that compensate (negative $\Gamma$) or add dispersion (positive $\Gamma$) to the intrinsic phase of the resonance.

Two quadrants in the phase-space, II and IV, exhibit clear enhancement in the SFG signal relative to the TL case. In quadrant IV, the enhancement peaks when the applied Atan phase compensates the resonant phase $\phi_E(\Omega,\Gamma)=-\phi_D(\omega_c+\Delta\omega_0, \gamma_0)$, yielding a temporally compressed oscillator displacement ($\phi_{x_0}(\omega_c+\Delta\omega_0, -\gamma_0)=const$) and maximal nonlinear response. In contrast, the enhancement in quadrant II originates from a different mechanism. Here, the applied spectral phase adds dispersion to the intrinsic resonance phase, resulting in an overall antisymmetric displacement phase profile around $\omega_c$, i.e., $\phi_{x_0}(-\Delta\omega_0, \gamma_0)=-\phi_{x_0}(\Delta\omega_0, \gamma_0)$ (ignoring global phases). 
Similar antisymmetric phases are known to preserve two-photon absorption efficiency in nonresonant excitations,\cite{meshulach1998} yet their effect in resonant scenarios has not been previously reported. The antisymmetric displacement phase profile ensures that photon pairs symmetric about the carrier frequency $\omega_c$ experience identical group delays when interacting with the resonance, i.e., $\tau_g(\omega_c + \Delta\omega) = \tau_g(\omega_c - \Delta\omega)$. This group delay symmetry preserves the phase relationships necessary for constructive interference, as frequency-symmetric photon combinations arrive simultaneously despite the dispersive resonant medium. Consequently, this antisymmetry leads to constructive interference among all two-photon pathways leading to $2\omega_c$, and manifests in spectral broadening of the SFG response (Fig.~\ref{fig3}d).
In the remaining quadrants (I and III), the spectral phase distorts this symmetry, resulting in destructive interference and suppressed nonlinear output, consistent with previously described 'dark pulse' conditions~\cite{silberberg2009}. This phase-space symmetry suggests a general design principle: tailoring the spectral phase to enforce antisymmetry (or group-delay symmetry) enables steering the system toward regimes of constructive multiphoton interference. While this understanding emerges from a second-order model, it offers predictive value for more complex higher-order processes, as shown in the FWM response of Fig.~\ref{fig2}.

Having demonstrated how tailored Atan spectral phases modulate FWM and SFG through coherent multiphoton interference, we extend our analysis to higher-order nonlinear processes. In particular, we investigate harmonic generation beyond third-order under resonant excitation, where the role of spectral phase remains largely unexplored. We restrict ourselves to classical perturbative simulations based on Eq.~\ref{eq1}, which models the oscillator dynamics in the weak-field regime. The relevance of resonantly enhanced harmonic generation was first highlighted by Kim et al.~\cite{kim2008}, who demonstrated that plasmonic field localization in gold nanoantennas can facilitate high-harmonic generation (HHG) using femtosecond pulses from a modest oscillator. 

\begin{figure}[ht]
    \centering
    \includegraphics[width=1\textwidth]{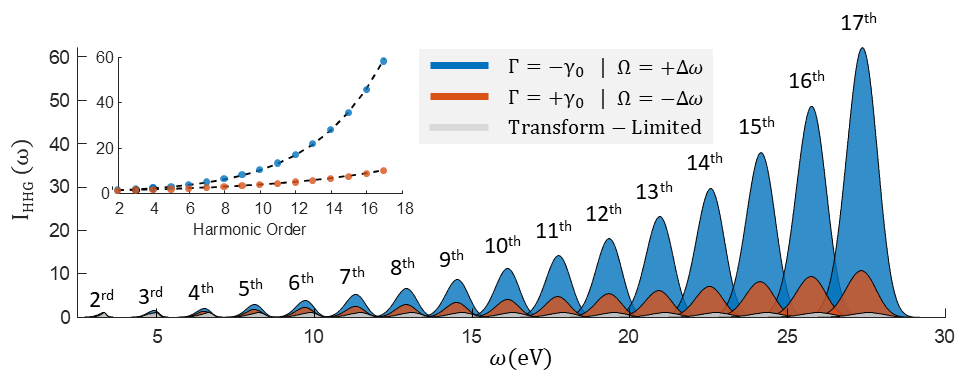}
    \caption{
Power spectral density of the HHG enhancement using two Atan-shaped pulses, normalized by the TL excitation response (gray area): An Atan phase compensating for the resonant phase (blue area), and an Atan phase creating an antisymmetric polarization response. Both phases show consistent enhancement and spectral broadening. 
Inset: the enhancement factor for both phases as a function of harmonic order, showing exponential scaling.}
    \label{fig4}
\end{figure}

Figure~\ref{fig4} illustrates the pronounced effect of Atan spectral phase shaping on HHG. Our simulations compare two distinct phase strategies, corresponding to quadrants II and IV in Fig.~\ref{fig3}. One compensates the intrinsic resonant phase (blue spectra), while the other introduces an antisymmetric displacement phase structure (orange spectra). Both approaches yield substantial enhancement relative to the TL case (gray spectra), with an enhancement factor that increases with harmonic order almost exponentially. 
For a 6\,fs driving pulse, the enhancement at the 17\textsuperscript{th} harmonic exceeds a factor of 58. In both cases, the harmonic spectra exhibit significant broadening. Notably, the antisymmetric displacement phase, despite not compensating the resonance-induced phase, enforces a symmetric group delay that enables constructive multiphoton interference. These results establish that both phase compensation and antisymmetric shaping offer distinct and effective routes for enhancing resonant nonlinear generation.

In conclusion, we have demonstrated a deterministic coherent control strategy for enhancing nonlinear optical processes in resonant plasmonic nanostructures using tailored Atan spectral phases. This approach uncovers a symmetric phase-space landscape of enhancement and suppression, governed by the interplay between spectral detuning and phase parity. Crucially, the underlying methodology is general and can be extended not only to arbitrarily shaped nanoparticles but to any resonant system that follows the classical anharmonic dynamics derived in our model.\cite{meron2025} Our simulations indicate that the enhancement factor rises with harmonic order, the resonance quality factor, and the bandwidth of the driving pulse. Notably, our sub-10\,fs single-pulse apparatus and phase-selective nonlinear measurements enable access to the near-field LSPR response approaching its homogeneous limit, effectively filtering out inhomogeneous broadening contributions that typically obscure resonant parameters in linear far-field spectroscopy.\cite{sonnichsen2002, lietard2018} Furthermore, the emergence of a nonintuitive secondary enhancement peak for an applied phase that modifies the LSPR polarization response to become antisymmetric improves nonlinear generation efficiency while sustaining low peak power for dispersed pulses. 
Looking forward, it would be particularly interesting to apply this 2D phase-space approach to more complex resonant systems, such as coupled nanoresonators,\cite{funston2009, liu2010, aeschlimann2016} Fano resonances,\cite{lukyanchuk2010, faggiani2017} BICs (Bound states in the Continuum),\cite{zograf2022} or dark-bright mode hybrids.\cite{zhang2008,liu2018} Using this method, we may be able to disentangle the rich interference phenomena in such systems and thereby unravel the underlying microscopic mechanisms driving their nonlinear responses.\cite{utikal2011, meron2025} 
 
\begin{acknowledgement}

We acknowledge the funding by the Israel Science Foundation (ISF) Grant No. 2312/21. 

\end{acknowledgement}

\begin{suppinfo}
SEM images and linear reflection spectrum fitting are presented in supplementary notes S1-S2, Figs. S1-S2.
\end{suppinfo}

\providecommand{\latin}[1]{#1}
\makeatletter
\providecommand{\doi}
  {\begingroup\let\do\@makeother\dospecials
  \catcode`\{=1 \catcode`\}=2 \doi@aux}
\providecommand{\doi@aux}[1]{\endgroup\texttt{#1}}
\makeatother
\providecommand*\mcitethebibliography{\thebibliography}
\csname @ifundefined\endcsname{endmcitethebibliography}  {\let\endmcitethebibliography\endthebibliography}{}

\end{document}